\newcommand{\Delq}{\Delta_{q}}
\newcommand{\tauq}{\tau(q)}
\newcommand{\taupq}{\tau'(q)}
\newcommand{\delq}{\delta_{q}}
\newcommand{\Delminq}{\Delta_{1-q}}

\newcommand{\fa}{f(\alpha)}
\newcommand{\fpa}{f'(\alpha)}
\newcommand{\psisq}{\psi^2}
\newcommand{\vecr}{\mathrm{\mathbf{r}}}

\documentclass[aps,prl,twocolumn,superscriptaddress,showpacs]{revtex4}
\usepackage{graphicx}
\begin{document}
%\linenumbers
%\preprint{version 2.31}
\title{Observation of multifractality in
Anderson localization of ultrasound}


\author{Sanli Faez}
\email[]{faez@amolf.nl}
%\thanks{}
\affiliation{FOM Institute for Atomic
and Molecular Physics AMOLF, Science
Park 113, 1098 SJ Amsterdam, The
Netherlands}
\author{Anatoliy Strybulevych}
\affiliation{Department of Physics and
Astronomy, University of Manitoba,
Winnipeg, Manitoba, R3T 2N2,
Canada}
\author{John H. Page}
\affiliation{Department of Physics and
Astronomy, University of Manitoba,
Winnipeg, Manitoba, R3T 2N2, Canada}
\author{Ad Lagendijk}
\affiliation{FOM Institute for Atomic
and Molecular Physics AMOLF, Science
Park 113, 1098 SJ Amsterdam, The
Netherlands}
\author{Bart A. van Tiggelen}
\affiliation{Universit\'{e} Joseph
Fourier, Laboratoire de Physique et
Mod\'{e}lisation des Milieux
Condens\'{e}s, CNRS, 25 Rue des
Martyrs, BP 166, 38042 Grenoble,
France}

\date{\today}

\begin{abstract}
We report the experimental
observation of strong multifractality
in wave functions below the Anderson
localization transition in open
three-dimensional elastic networks. Our
results confirm the recently predicted
symmetry of the multifractal exponents.
We have discovered that the result of
multifractal analysis of real data
depends on the excitation scheme used in
the experiment.
\end{abstract}

%72.15.Rn   Localization effects (Anderson or weak localization)
%05.70.Jk   Critical point phenomena
%42.25.Dd   Wave propagation in random media
%71.30.+h   Metal–insulator transitions and other electronic transitions
%\keywords{}

\maketitle
%\section{\label{sec:Intro}Introduction}
Critical phenomena are of prominent
importance in condensed-matter physics.
Criticality at the Anderson
localization transition has been the
subject of intensive theoretical
research.
Some important theoretical predictions
have been made, among which is the remarkable aspect of
multifractality of wave functions at this transition.
Numerical simulations support these predictions
but also raise more questions~\cite{evers_anderson_2008}.
Recent experimental progress has paved the way
for the direct investigation of the
Anderson localization transition at the mobility edge in real
samples~\cite{morgenstern_real-space_2003,hu_localization_2008,chabe_experimental_2008}.

In this Letter, we report the
experimental observation of strong multifractality
(MF) just below the Anderson transition. This
observation is based on the excitation of
elastic waves in an open 3D disordered
medium. The recently predicted symmetry
relation of the MF
exponents~\cite{mirlin_exact_2006} is
tested and confirmed. All results are
compared with the corresponding
analysis of diffusive (metallic) wave
functions in the same network at a
different frequency or with a light
speckle pattern generated by a strongly
scattering medium, showing a very clear difference between
localizing and diffusive regimes. Our
results not only highlight the
presence of MF in wave functions close to the
mobility edge, but also reveal new
aspects of the MF character in real experimental
systems.

Before presenting the experimental
results, we briefly review some general
aspects of MF and their implications in
the context of the Anderson transition.
Multifractality quantifies
the strong fluctuations
of the wave function.
It shows the non-trivial
length-scale dependence of the moments
of the intensity distribution. The
dependence can be investigated by
varying the system size $L$, or
alternatively, if the system size is
fixed, by dividing the system into
small boxes of linear size $b$ and
varying $b$. This property is
quantified by using the generalized
Inverse Participation Ratios (gIPR)
\begin{equation}\label{eq:definitionipr}
   P_q = \sum_{i=1}^n \left(I_{B_i}\right)^q=\sum_{i=1}^n \left[\int_{B_i} {I(\vecr)} d^d \vecr\right]^q,
\end{equation}
where $I(\vecr)$ is the normalized
intensity (equal to
$|\psisq(\vecr)|/\int{|\psisq(\vecr)|d^dr}$
where $\psi(\vecr)$ is the wave
function) and $I_{B_i}$ is the
integrated probability inside a box
$B_i$ of linear size $b$, with
$\lambda\ll b\ll L$ where $\lambda$ is
the wavelength. The summation is
performed on the whole sample, which
consists of $n=(L/b)^d$ boxes, and $d$
is the space dimension. By definition
$P_1\equiv1$ and $P_0\equiv n$.

At criticality, the ensemble averaged
gIPR, $\langle P_q\rangle$, scales
anomalously with the dimensionless
scaling length $L/b$ as
\begin{equation}\label{eq:iprscaling}
   \langle P_q \rangle \sim
   \left({L}/{b}\right)^{-d(q-1)-\Delq}\equiv \left({L}/{b}\right)^{-\tauq},
\end{equation}
where $d(q-1)$ and $\Delq$ are called
the normal (Euclidean) and the
anomalous dimensions, respectively. For
a normal (extended) wave function,
$\Delq = 0$ for every $q$. A (single-)
fractal wave function with fractal
dimension $D$ is described by
$\tauq\equiv D(q-1)$. For critical
states $\tauq$ is a continuous function
of $q$ that fully describes the MF.

MF describes the scaling of the moments
of a probability density function
(PDF). The gIPR, defined in
Eq.~(\ref{eq:definitionipr}), are
proportional to the moments of the
distribution function of the
eigenfunction intensities, so that Eq.~(\ref{eq:iprscaling})
implies the following
scaling relation for the PDF:
\begin{equation}\label{eq:pdfandmf}
   \mathcal{P}(\ln I_B)\sim \left({L}/{b}\right)^{-d+f\left(-\frac{\ln I_B}{\ln \frac{L}{b}}\right)}.
\end{equation}
The second term in the exponent, $\fa$,
is called the singularity spectrum, and
is related to the set of anomalous
exponents $\tauq$ by a Legendre
transform
\begin{equation}\label{eq:ltransform}
\tauq=q\alpha-\fa,\;\;\;q=\fpa,\;\;\;\;\alpha=\taupq.
\end{equation}

The singularity spectrum $\fa$ is the
fractal dimension of the set of those
points $\vecr$ where the wave-function
intensity, $I(\vecr)$, scales as
$L^{-\alpha}$. In mathematical terms,
it shows the coexistence of several
populations of singularities in the
measure, which is the wave-function
intensity for this specific case. In
the field-theoretical treatment of
random-Schr\"{o}dinger Hamiltonians, MF
implies the presence of infinitely many
relevant
operators~\cite{Ludwig_PRL1991,
Mudry_1996}. The functional dependence
of $\fa$ is an important and unique
property of each universality class. In
the extended regime, $\mathcal{P}(\ln
I_B)$ is strongly peaked near
$\alpha=d$, since the short-range
``Gaussian''
fluctuations~\cite{Shapiro_PRL86} are
washed out in the box integration.

First order perturbation theory for an
Anderson transition in
$2+\epsilon$~dimensions~\cite{wegner_inverse_1980} (two is the
critical dimension), results in the
``parabolic approximation'' for the MF
wave
functions~\cite{aoki_critical_1983,castellani_multifractal_1986}.
This result,
$\Delq=\gamma q(1-q)$, corresponds to
$\fa=d-(\alpha-d-\gamma)^2/4\gamma$,
where $\gamma$ is a constant in the order of $\epsilon$. A
similar approximation applies to metallic (diffusive) states
in three dimensions~\cite{altshuler-JETP86,falko_statistics_1995}
due to weak localization, although with $\gamma \ll
1$. This is sometimes called weak MF.

Recently, an exact symmetry relation
\begin{equation}\label{eq:symmetry}
\Delq=\Delminq,
\end{equation}
was theoretically predicted for the set
of anomalous exponents~\cite{mirlin_exact_2006}.
The numerical and analytical
investigations of the 3D Anderson model and
certain random matrices suggest that MF
may exhibit itself also for off-criticial
states on both sides of the transition~\cite{cuevas_prb_2007}.
The MF concept was extended to the
boundaries where it behaves differently
with respect to the
bulk~\cite{subramaniam_surface_2006}.

Most of the available information about
MF is based on numerical
investigations (See, e. g.,~\cite{evers_anderson_2008,grussbach_determination_1995,mildenberger_wave_2007,Ludlam_PRB_2003} and references therein).
The only experimental attempt to
observe strong MF in wave functions so
far is due to Morgenstern~\emph{et.~al}
using scanning tunneling microscopy of
2D electron
systems~\cite{morgenstern_real-space_2003}.
Their observation of MF was hindered by
the presence of several eigenfunctions
in the measurement and by the limited
size of their system.
%\section{\label{sec:experiment}Experiment}

We now use ultrasonic measurements to
demonstrate three different, but
closely related, manifestations of MF:
(1) the probability density function
(PDF), (2) the scaling of generalized
Inverse Participation Ratios (gIPR),
and (3) direct extraction of the
singularity spectrum. Our experiments
were performed on disordered single-component
elastic networks, made by
brazing randomly-packed $4.11$-mm-diameter
aluminium beads together~\cite{hu_localization_2008}. The
data presented here were obtained from
a representative disc-shaped sample
with a 120~mm diameter and 14.5~mm
thickness. Two different configurations
were used for excitation. In the first
excitation scheme a point-like
ultrasound source emits short pulses
next to the sample surface. In the
second case the source was put far from
the sample so that a quasi-planar wave
was incident on the whole interface. In
close proximity to the opposite
interface, vibrational excitations of
the network were probed with
sub-wavelength-diameter detectors in
the frequency range of 0.2 to 3~MHz, where
the wavelengths are comparable to the bead size
and the scattering mean free paths are much less than
the sample thickness~\cite{hu_localization_2008}.
The intensity at a particular frequency
was determined from the square of the
magnitude of the Fourier transform of
the entire time-dependent transmitted
field in each near-field speckle. The
intensity was normalized by the total
intensity in the measured speckle
pattern. The normalized speckle
intensity, $I(j)$ was recorded at each
point $j$ on a square grid of linear
size $L_g=55$ points with a typical
nearest-neighbor spacing of 0.66~mm.

In the lower frequency band around
250~kHz, the ultrasound propagation is
diffusive. A localizing regime is
observed in a 50\% bandwidth around
2.4~MHz, where the measured localization length in
the sample is smaller than the size of the
analyzed speckle patterns ($0.7 L_g$) and almost equal
to the sample thickness. A full description of the
experiment and a thorough comparison of
previous measurements with the self-consistent theory of
localization has been presented
in~\cite{hu_localization_2008}.

%The
%results presented in this Letter are
%based mostly on new experiments using
%using point-source excitation, and involve entirely
%different methods of analysis.

%\section{\label{sec:analysis}Analysis}
\begin{figure}
\includegraphics[width=6.4cm]{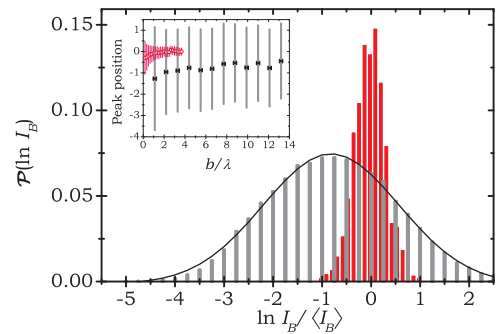}%
\caption{\label{fig:distribution}Comparison
between coarse-grained PDF for
localized and diffusive speckle
intensities. The PDFs are
experimentally obtained from the
histogram of the logarithm of averaged
intensities in the localized (thick
bars) and diffusive (thin bars)
regimes. The black line shows a fit to
a single parameter log-normal
distribution given by the parabolic
approximation of Eq.~\ref{eq:pdfandmf}.
Inset: The peak position (symbols) and
the full width at half maximum (bars)
of the intensity histogram is plotted
for localized (circles) and diffusive
(squares) speckles as a function of
coarse-graining box size.}
\end{figure}

We obtain the PDF from the histogram of
the logarithm of box-integrated
intensities $I_{B_i}$. We sample over
100 speckles in a 5\% bandwidth around
250~kHz and 2.4~MHz for diffusive and
localized regimes, respectively. Two
representative histograms are shown in
Fig.~\ref{fig:distribution} with
typical box sizes of $b=9$ and $b=2$
points
for low and high frequency
measurements,
corresponding to box sizes of
approximately two wavelengths in both
cases. The PDF for localized waves is
clearly much wider than the one for
diffusive waves and the peak is shifted
from the average intensity. We have
also plotted the peak-position and the
width of the histogram as a function of
box size in the inset of
Fig.~\ref{fig:distribution}.

\begin{figure}
\includegraphics[width=8.1cm]{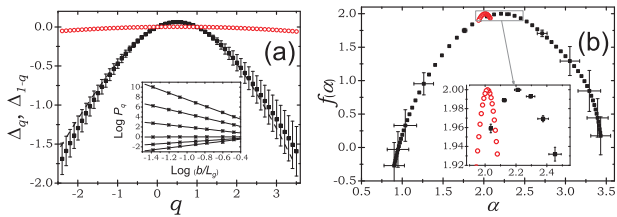}
\caption{\label{fig:moments} (a) The
measured anomalous exponents $\Delq$
are shown for localized ultrasound
(full squares) and diffusive light
(open circles) speckles. The dashed
line shows the same data-points,
mirrored relative to $q=\frac{1}{2}$ in
order to check the symmetry in the
spectrum. The anomalous exponents are
estimated from the box-counting method.
The slope of the gIPR plotted versus
the box size in bilogarithmic scales
yields $\Delq$. One example is shown in
the inset for $q\in\{-2,-1,0,1,2,3\}$
and $f=2.40$~MHz. (b) The average
singularity spectrum is calculated for
the ultrasound speckles (full squares)
at frequencies between 2.0 to 2.6 MHz.
For comparison a singularity spectrum
for diffusive optical speckle (open
circles), with the Euclidian dimension,
is extracted by applying the same
box-counting procedure.}
\end{figure}

In principle, it is possible to extract
the MF spectrum from the
PDF~\cite{rodriguez_multifractal_2009}.
However, a box-counting analysis can
give more accurate results based on the
scaling of the gIPR. Similar to many
numerical studies, we approximate the
expectation values by box-sampling over
a single or multiple wave functions
measured for a single realization of
disorder. This approximation is known
as typical averaging. Typical averaging
is unable to reveal information that is
related to statistically rare
events~\cite{evers_anderson_2008}. In
this approach, the system size is fixed
and supposed to be large enough
relative to the box size. The
approximate scaling relation is derived
by plotting the estimated gIPR, given
by Eq.~(\ref{eq:definitionipr}), versus
the box size $b$~\footnote{We have used
box sizes $b\in\{2,3,4,6,8,12,24\}$}.
Note that although we have examined
three-dimensional samples, the
Euclidian dimension of our sampling
space is two since the available data
is taken just from the surface of the sample.
The effective system size is $L_g$ over
which the intensities are normalized.
By plotting ${P_q}$ versus the box size
in bilogarithmic scales [e.g., see the
inset to Fig.~\ref{fig:moments}(a)],
power-law behavior is found for $q
\in[-3,4]$, with the slope yielding the
scaling exponent $\tau(q)$. The average
anomalous exponent is obtained by
averaging the exponents measured for
several frequencies between 2.0 and
2.6~MHz and subtracting off the normal
part of the exponent $2(q-1)$. The
standard deviation is taken as the
error-bar.

The anomalous exponents are plotted as
a function of $q$ in
Fig.~\ref{fig:moments}(a). For
comparison with the localized data, the same numerical
procedure was applied to a
diffusive speckle pattern, where the behavior
is entirely different ($\Delq=0$).
In making this comparison, an optical
diffusive speckle pattern was used to
capitalize on the best available
statistics.

The behavior of the anomalous exponents
shown in Fig.~\ref{fig:moments}
provides unambiguous evidence for surface MF of
the localized ultrasound wave
functions.  This is the most important
result in this Letter.
Note that MF is clearly seen in these data,
even though the localized wave functions in
our finite sample are near to, but not exactly at, criticality.
In addition, our
observation of MF clearly supports the
predicted symmetry
relation~(\ref{eq:symmetry}).  Our
experimental demonstration of this
fundamental symmetry, seen in a very
different system to the ones envisaged
in~\cite{mirlin_exact_2006}, attests to
the universality of critical properties
near the Anderson transition.

Finally, we have extracted the surface
MF spectrum directly from the
measurements~\cite{chhabra_direct_1989}.
In this method the numerical error
caused by the Legendre
transform~(\ref{eq:ltransform}) is avoided. To get
enough statistics, 100 wave functions
in a bandwidth of 5\% are used to
estimate the MF spectrum for several
seven frequency bands between 2.0 and
2.6~MHz. No systematic deviation is
observed between the seven spectra
obtained in this frequency range. These
spectra are then averaged for each
value of $q\in[-6,6]$ and the standard
deviation is considered as the error
bar. The results are summarized in
Fig.~\ref{fig:moments}(b). The peak of
the MF spectrum is shifted from two
(the Euclidian dimension of the
measurement basis) by a value of
$0.21\pm0.02$. For comparison, the same
procedure is applied to the optical
speckle using the same $q$-range. No
shift is observed for the optical
speckle.

\begin{figure}
\includegraphics[width=8.1cm]{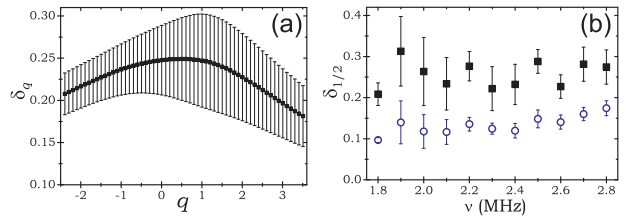}
\caption{\label{fig:parabolic}(a) The
reduced anomalous dimension
$\delq\equiv\frac{\Delq}{q(1-q)}$ is
plotted versus $q$. Bars show the
estimated error. Deviation from a
horizontal line corresponds to the
deviation from parabolic approximation.
(b) The reduced anomalous dimension
$\delta_{\frac{1}{2}}$ is plotted
versus frequency in the localized
regime for two excitation schemes:
point-source (squares) and plane wave
(circles). The error bars represent the
standard deviation of the measured
exponents that are averaged over each
0.1-MHz-wide frequency band.}
\end{figure}

The MF that is clearly seen in our data
allows us to test the deviation from
the parabolic approximation. This is
characterized by the reduced anomalous
exponents
$\delq\equiv\frac{\Delq}{q(1-q)}$. In
our results, shown in
Fig.~\ref{fig:parabolic}(a), we see a
deviation of less than 20\% for
$q\in[-3,4]$. The non-parabolicity of
the spectrum is very difficult to measure but it may have
important theoretical consequences.
More precise investigation of larger samples is
needed to reliably confirm or exclude
the possibility of a small but significant deviation.

%\section{discussion}
We have also investigated the
dependence of the reduced anomalous
exponent at the symmetry axis,
$\delta_{\frac{1}{2}}=4\Delta_{\frac{1}{2}}$,
on the frequency and type of
excitation. The results are presented
in Fig.~\ref{fig:parabolic}(b). We
observe a robust presence of MF for all
frequencies between 1.7 to 2.9~MHz. The
measured anomalous exponent is larger
for the point source illumination. This
difference may be related to the number
of modes excited in each scheme. It has
been previously
discussed~\cite{terao_multifractality_1996}
that the overlap of two or more
eigenmodes shifts the peak of the
singularity spectrum towards the
Euclidian dimension. Since the surface area of the sample
is larger than the localization length, neighboring
localized modes may coexist at the same
frequency. These modes can all be
excited by a quasi-plane wave while a
point source couples more efficiently
to the closest mode.

%\subsection{\label{subsec:deviation}Deviation from numerical results}
Numerical analyses of bulk and surface MF
for the eigenstates of the Anderson tight-binding Hamiltonian
on a 3D cubic lattice at the metal-insulator transition
have predicted corresponding shifts
of $1.0$ and $1.6$ from the Euclidean dimension
for the peak of the singularity
spectrum~\cite{monthus_surface_2009,rodriguez_multifractal_2009}.
Another numerical study for an equivalent vibrational model on
the \emph{fcc} lattice shows a
similar outcome for bulk MF~\cite{Ludlam_PRB_2003},
indicating to the universality of this
phenomenon. It is not simple to explain
the difference between the available
numerical results and our experimental
outcome. Several issues may play a
role. Mode overlap and the finite
lifetime of modes due to open
boundaries are two of these issues. Most numerical
studies are done based on uncorrelated
disorder, which is experimentally hardly ever
achieved. The uniform bead size in our samples, which is comparable with the vibrational
wavelength, is a source of correlation.
The presence of correlation in
the disordered potential may influence
the critical behavior and
induce nonuniversality~\cite{demoura_correlated_2007}.

%\subsection{\label{subsec:remarks}Final remarks}
Despite the wealth of theoretical and
numerical studies on the Anderson
transition in 2D and 3D for the
Schr\"{o}dinger Hamiltonian in closed
systems, critical properties of this transition for classical waves
in an open system have never been
studied. Our system is specially
challenging due to its 3D nature, open
boundaries and presence of three
polarizations for the elastic waves.
Specific properties of classical waves
such as absorption are yet to be
investigated in the context of
criticality. Our results show that these
important questions can now be investigated
experimentally, providing vital guidance for new
theoretical work. Our experiments reveal
that the concept of MF not only concerns
critical states but is valid as well around the mobility
edge. This observation agrees with recent theoretical investigations~\cite{cuevas_prb_2007}.
Mutual avoidance of wave functions at
large energy separations and their enhanced
overlap at small energy separations are other important predictions of that
theory, which can also be verified by our experiment.

In conclusion, we have presented the
first experimental observation of
multifractal wave functions below the
localization transition. Our
data validate experimentally the
predicted symmetry relation of the
anomalous exponents. Free from
interactions and with the possibility
of diverse illumination and detection
schemes, sound and light experiments
can provide a tremendous amount of
useful information for this field of
research. We believe that our
observation of multifractality in
classical waves will stimulate new
theoretical and numerical
investigations. On the experimental
side, this work highlights again the
strength of statistical methods for
studying Anderson localization.

%\appendix
%\section{}

\begin{acknowledgments}
We thank J. S. Caux, R. G. S.
El-Dardiry, F. Evers, P. M. Johnson, A.
D. Mirlin, and S. E. Skipetrov for
discussions and H. Hu for making the
samples. Financial support by the
``Nederlandse Organisatie voor
Wetenschappelijk Onderzoek'' (NWO), the
``Centre National de la Recherche
Scientifique'' (PICS), and the
``Natural Sciences and Engineering
Research Council of Canada'' (NSERC) is
acknowledged.
\end{acknowledgments}

%\bibliography{mflibrary2}

\end{document}